\documentclass[preprint,aps,prc,showpacs,nofootinbib]{revtex4}
\usepackage{epsfig}
\usepackage{graphicx}

\newcommand{\Li}{\mbox{Li}_2}

\newcommand{\Ree}{\mbox{Re}}
\newcommand{\dd}{\mbox{d}}
\newcommand\ba{\begin{eqnarray}}
\newcommand\ea{\end{eqnarray}}
\newcommand\nn{\nonumber}
\def\Li#1#2{{\mathrm{Li}}_{#1}\left(#2\right)}

\begin{document}
\title{Target normal spin asymmetry and charge asymmetry for $e\mu$  elastic
scattering and the crossed processes}

\author{E. A. Kuraev, V. V. Bytev, Yu. M. Bystritskiy}
\affiliation{\it JINR-BLTP, 141980 Dubna, Moscow region, Russian
Federation}

\author{E. Tomasi-Gustafsson}

\affiliation{\it DAPNIA/SPhN, CEA/Saclay, 91191 Gif-sur-Yvette
Cedex, France }

\date{\today}

\begin{abstract}
Two kinds of asymmetry arise from the interference of the Born amplitude
and the box-type amplitude corresponding to two virtual photons exchange, namely
charge-odd and one spin asymmetries. In case of unpolarized particles the charge-odd
correlation is calculated. It can be measured in combination of electron muon and positron
muon scattering experiments. The forward-backward asymmetry is the corresponding quantity
which can be measured for the crossed processes. In the case of polarized muon  the one-spin
asymmetry for annihilation and scattering channels has been calculated. The additional structure
function arising from the interference is shown to suffer from infrared divergencies.
The background due to electroweak interaction is discussed.
\end{abstract}

\maketitle

\section{Introduction}
The motivation of this work is to give an accurate description of the process $e\bar\mu\to e\bar\mu(\gamma)$
$e\bar{e}\to \mu\bar\mu(\gamma)$  in frame of QED, in order to provide a basis for the comparison with experimental data. High precision experiments on the processes $e\bar{e}\to\tau\bar\tau$ and $e\bar{e}\to p\bar{p}$ are planned in future $c-\tau$ facilities.  Moreover, the possibility of colliding $e\mu$ beam facilities has been discussed in framewok of programs on verification of Standard Model (SM) prediction.

The obtained results can also be applied, as a realistic model, to electron
(positron) scattering on a point-like hadron (proton), based on arguments given in this paper.

It is known  that in Born approximation the differential cross-section of
elastic proton-electron scattering
\begin{equation}
e(p_{1}) + p(p) \to e(p_{1}') + p(p') \label{eq:eq1}
\end{equation}
can be expressed in terms of two proton form factors,
$F_{1,2}(q^2)$, which are functions of a single argument, the  momentum transfer squared, $q^2=t$.

Taking into account two (and more) photon exchanges, (TPE), leads to a  generalization of the Born picture, namely the amplitude of  $ep$ scattering depends on two Mandelstam variables, the total energy  $s$ and $t$. The virtual photon Compton scattering amplitude is a rather complex object,
which can be expressed in terms of 12 chiral amplitudes. Nevertheless, taking into account parity conservation and omitting the terms of order $m_e/m_\mu$, (which are responsible
for chirality violation) reduce the number of relevant amplitudes to three \cite{Gold57}:
\begin{eqnarray}
\nonumber M^{(2)} =\frac{ i\alpha^2}{t} \bar{u}(p_{1}') \gamma_{\mu} u(p_{1})\times
\bar{u}(p')
\biggl[F_{1}(s,t)\gamma_{\mu}-\frac{F_{2}(s,t)}{2M}\gamma_{\mu}\hat{q}+\frac{1}{t}F_{3}(s,t)(\hat{p}_1+\hat{p}_1')
(p+p')_\mu\biggr] u(p),
 \label{eq:eq2}
\end{eqnarray}
with $q=p_{1}-p_{1}'$, $s=2p_{1}p$.

The explicit calculation, given in this paper, permits one to extract the individual contributions $F_1$, $F_2$, and $F_3$, in frame of QED. The infrared divergency is cancelled when the relevant soft photon emission is correctly taken into account.

Charge-odd and backward-forward asymmetries appear naturally
from the interference of one and two photon exchange amplitudes in frame of QED and SM due to $Z_0$-boson exchange in Born approximation. But at  the energy range reachable at $c-\tau$ factories, the relevant contribution
of SM type is \cite{VanNieu}:
\ba
\frac{\dd\sigma^{odd}_Z}{\dd\sigma_{QED}}\approx \frac{s}{M_Z^2}a_va_a\approx5\cdot10^{-5},
\quad 3<\sqrt{s} <5~\mbox{GeV}.
\ea
which is quite small compared to QED effects.

The accuracy of results given below is determined  by
\ba
\mathcal{O}\biggl(\frac{m_e^2}{m_\mu^2},\frac{m_e^2}{m_\tau^2},\frac{m_e^2}{m_p^2}\biggr)
\sim 0.1 \%
\label{eq:eq9}
\ea
and the contribution of higher orders of QED $\alpha/\pi\approx 0.5$\%.
Moreover we assume that all the velocities of the final heavy particles are finite in the annihilation as well as in the scattering channels.
This is the reason why Coulomb factors are neglected.

Our paper is organized as follows. In sections \ref{ChargeOddAsymmetries} the annihilation channel and the scattering channel $e\bar{e}\to \mu\bar\mu$ are considered in sections \ref{ChargeOddAsymmetries} and \ref{scatt_channel}, respectively.
In section \ref{soft_phot_glav} we take into account the soft photon emission and construct charge-odd
and forward-backward asymmetries.
In section \ref{CrossingSymmetry} we analyze the crossing relation between two channels.
Explicit form for additional structure $G_3$ for annihilation channel is given in section \ref{DerivationOfF3}.
In section \ref{ProtonSpinAsymmetry} the one-spin asymmetries are investigated.
The results are summarized in the Conclusions.

\section{Process $\lowercase{e^++{e}^-\to \mu^++{\mu}^-(\gamma)}$}
\label{ChargeOddAsymmetries}

At first, we consider the process of creation of ${\mu}^+\mu^-$ pairs in
electron-positron annihilation:
\begin{equation}
e^{+}(p_{+}) + e^{-}(p_{-}) \to \mu^{+}(q_{+}) + \mu^{-}(q_{-}).
\label{eq:eq3}
\end{equation}
The cross-section in the Born approximation, can be written as:
\begin{equation}
\frac{d\sigma_B}{dO_{\mu_{-}}} = \frac{\alpha^{2}}{4s} \beta
(2-\beta^{2}+\beta^{2} c^2), \label{eq:eq4}
\end{equation}
with $s=(p_{+}+p_{-})^{2}=4E^2$, $\beta^{2}=1-\frac{4m^{2}}{s}$,
 $E$ is the electron beam energy in center of mass reference
frame (implied for this process below), $m,m_e$ are the masses of muon and electron,
$c=\cos\theta$, and $\theta$ is the angle of $\mu_{-}$-meson emission to the electron beam direction.

The interference of the Born amplitude
\begin{eqnarray}
\nonumber M_{B} = \frac{i4\pi\alpha}{s} \bar{v}(p_{+}) \gamma_{\mu}
u(p_{-}) \bar{u}(q_{-}) \gamma_{\mu} v(q_{+}),
\label{eq:eq5}
\end{eqnarray}
with the box-type amplitude $M_B$, results in parity violating contributions to the differential cross section,
i. e. the ones, changing the sign
at $\theta \to \pi-\theta,$. As a consequence  of
charge-odd correlations  we can construct:
\begin{equation}
A(\theta,\Delta E) = \frac{\dd\sigma(\theta)-\dd\sigma(\pi-\theta)}{\dd\sigma_B{(\theta)}}.
\label{eq:eq6}
\end{equation}
Here we take into account as well the emission of an additional soft real photon
with energy not exceeding some small value $\Delta E$, so that $A(\theta,\Delta E)$
is free from the infrared singularities.

Part of the results presented here were previously derived by one of us (E. A. K.) in  Ref. \cite{KM76a}, and partially published in \cite{KM76b}.

\begin{figure}
\begin{center}
\includegraphics[width=12cm]{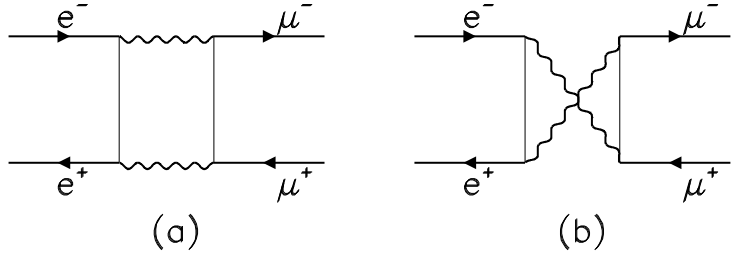}
\caption{\label{Fig:fig1} Feynman diagrams for two-photon
exchange in $e\bar{e}\to\mu\bar{\mu}$ process: box diagram (a) and
crossed box diagram (b).}
\end{center}
\end{figure}

There are two box-type Feynman amplitudes (Fig. \ref{Fig:fig1}). We calculate only one of
them, the uncrossed diagram (Fig. \ref{Fig:fig1}a) with matrix element
\ba
&M_{a} &= i\alpha^2\int \frac{\dd^{4}k}{i\pi^{2}} \frac{\bar{u}(q_{-})T v(q_{+})\times
\bar{v}(p_{+})Z
u(p_{-})}{(\Delta)(Q)(P_+)(P_-)},
\nonumber \\
&&(\Delta)=(k-\Delta)^{2}-m_{e}^{2},\quad (Q)=(k-Q)^{2}-m^{2},\quad (P_\pm)=(k\mp P)^{2}-\lambda^{2},
\label{eq:eq8}
\ea
with $\lambda$-"photon" mass and
\ba
&T&=\gamma_{\alpha}(\hat{k}-\hat{Q}+m)\gamma_{\beta}, \quad \nonumber
Z=\gamma_{\beta}(\hat{k}-\hat{\Delta})\gamma_{\alpha}, \\
&&\Delta =\frac{1}{2}(p_{+}-p_{-}),\quad  Q= \frac{1}{2}(q_{+}-q_{-}),\quad
P=\frac{1}{2}(p_{+}+p_{-}).
\label{eq:design}
\ea
We will assume
\begin{equation}
m^{2}=\frac{s}{4}(1-\beta^2)\sim s\sim -t\sim -u. \label{eq:eq10}
\end{equation}

The explicit form of kinematical variables used below is:
\ba
\Delta^2=-P^2=-\frac{s}{4}, \quad  Q^2=-\frac{1}{4}s\beta^2,\quad \sigma=\Delta Q=\frac{1}{4}(u-t),
\nonumber \\
\quad u=(p_--q_+)^2=-\frac{s}{4}(1+\beta^2+2\beta c),
\quad t=(p_--q_-)^2=-\frac{s}{4}(1+\beta^2-2\beta c ).
\label{eq:defsu}
\ea
The contribution to the cross section of the amplitude arising from the crossed
Feynman diagram (Fig. \ref{Fig:fig1}b), $M_{b}$, can be obtained from $M_{a}$ by the crossing relation
\begin{equation}
\frac{\dd\sigma_{a}(s,t)}{\dd\Omega_{\mu}} =
-\frac{\dd\sigma_{b}(s,u)}{\dd\Omega_{\mu}}, \label{eq:eq11}
\end{equation}
which has the form
\begin{equation}
\frac{\dd\sigma_{a}(s,t)}{\dd\Omega_{\mu}} = \frac{\beta\alpha^{3}}{2\pi
s^{2}} \Ree[ R(s,t)],
\label{eq:eq12}
\end{equation}
with
\begin{equation}
R(s,t)=\int\frac{\dd^4k}{i\pi^2}\frac{1}{(\Delta)(Q)(P_+)(P_-)}\frac{1}{4}
Tr\left [\left ( \hat{q}_{-}+m)T (\hat{q}_{+}-m\right )\gamma_{\mu}\right ]
\times
\frac{1}{4}Tr\left(\hat{p}_{+}Z\hat{p}_{-}\gamma_{\mu}\right).
\label{eq:eq13}
\end{equation}
The scalar, vector and tensor loop momentum integrals are defined as:
\begin{equation}
J;J_{\mu}; J_{\mu\nu}=\int\frac{\dd^{4}k}{i\pi^{2}}
\frac{1;k_{\mu};k_{\mu}k_{\nu}}{(\Delta)(Q)(P_+)(P_-)} \label{eq:eq14}
\end{equation}
Using symmetry properties, the vector and tensor integrals can be written as:
\begin{equation}
J_{\mu}=J_{\Delta}\cdot \Delta_{\mu}+J_{Q}\cdot Q_{\mu},\label{eq:eq15}
\end{equation}
\begin{equation}
J_{\mu\nu}=K_{0}g_{\mu\nu}+K_{P}P_{\mu}P_{\nu}+K_{Q}Q^{\mu}Q^{\nu}
+K_{\Delta}\Delta_{\mu}\Delta_{\nu}+K_{x}(Q_{\mu}\Delta_{\nu}+Q_{\nu}\Delta_{\mu}).
\label{eq:eq16}
\end{equation}
The quantity $R(s,t)$ can be expressed as a function of polynomials $P_{i}$ as:
\begin{equation}
R=P_{1}J + P_{2}J_{\Delta} +P_{3}J_{Q}+P_{4}K_{0}
+P_{5}K_{\Delta}+P_{6}K_{Q}+P_{7}K_{P}+P_{8}K_{x},
\label{eq:eq17}
\end{equation}
where the explicit form of polynomials is given in Appendix A. Using the explicit
expression for the coefficients $J_{\Delta},...,K_{x}$ (See
Appendix B) we obtain
\begin{eqnarray}
\nonumber R(s,t)&=&4(\sigma-\Delta^{2})(2\sigma-m^{2})F
+16(\sigma-\Delta^{2})(\sigma^{2}+(\Delta^2)^2-m^{2}\Delta^{2})J \nonumber \\
&&+4[(\Delta^2)^2 -3 \Delta^{2}\sigma +2\sigma^{2}-m^{2}\sigma]F_{Q}
+ 4 [2(\Delta^2)^2 -2 \Delta^{2}\sigma
+2\sigma^{2}-m^{2}\Delta^{2}]F_{\Delta}
\nonumber\\
&&+4[(\Delta^2)^2+ \Delta^{2}\sigma+m^{2}\Delta^{2}]G_{Q}
+4[-(\Delta^2)^2+\sigma^{2}-2m^{2}\Delta^{2}]H_{Q},
\label{eq:eq18}
\end{eqnarray}
with the quantities $F\div H_{Q}$ given in Appendix B. Finally the
charge-odd part of differential cross section has the form
\ba
\left (\frac{\dd\sigma^{e\bar{e}}_{virt}(s,t)}{\dd\Omega_{\mu}}\right)_{odd} &=&
-\frac{\alpha^{3}\beta}{2\pi s} \mathcal{D}^{ann},
\nonumber \\
 \mathcal{D}^{ann}
&=&\frac{1}{s}[R(s,t)-R(s,u)]= (2-\beta^{2}+\beta^{2}
c^{2})\ln\left(\frac{1+\beta c}{1-\beta
c}\right)\ln\frac{s}{\lambda^{2}}+\mathcal{D}^{ann}_{V},
\label{eq:eq19}
\ea
\begin{eqnarray}
\nonumber
\mathcal{D}^{ann}_{V}&=&\left(1-2\beta^{2}+\beta^{2}c^{2}\right )
\left [
\frac{1}{1+\beta^{2}+2\beta c}\left( \ln\frac{1+\beta c}{2}+\ln\frac{s}{m^2}\right )
\right .
\\
&-&\left. \frac{1}{1+\beta^{2}-2\beta c}\left(\ln\frac{1-\beta c}{2}+\ln\frac{s}{m^2} \right )\right ]
\nonumber \\
&+& \beta c \left [
\phi(\beta)\left (\frac{1}{2\beta^{2}}-1-\frac{\beta^{2}}{2}\right )
-\frac{1}{\beta^{2}} \ln \frac{s}{m^{2}}
-\frac{\pi^{2}}{6} +\frac{1}{2} \ln^{2} \frac{s}{m^{2}}\right .
\nonumber \\
&-&\left. \frac{1}{2}\ln^{2}\frac{1-\beta c}{2}
-\frac{1}{2}\ln^{2}\frac{1+\beta c}{2}
+\Li{2}{\frac{1+\beta^{2}+2\beta c}{2(1+\beta c)}}
+\Li{2}{\frac{1+\beta^{2}-2\beta c}{2(1-\beta c)}}\right ]  \nonumber \\
&-&\frac{m^{2}}{s}\left [\ln^{2}\frac{1-\beta c}{2}
-\ln^{2} \frac{1+\beta c}{2}
+2\Li{2}{\frac{1+\beta^{2}+2\beta c}{2(1+\beta c)}} 
-2\Li{2}{\frac{1+\beta^{2}-2\beta c}{2(1-\beta c)}}
\right ],
\nonumber
\end{eqnarray}
where $\phi(\beta)=sF_Q$, $F_Q$ is given in Appendix B and
\ba
\Li{2}{z}=-\int\limits_{0}^{z}\frac{\dd x}{x}\ln(1-x)
\ea
is the Spence function. The quantity $\mathcal{D}^{ann} - \mathcal{D}^{ann}_{V}$ suffers from infrared divergences, which will be compensated taking into account the soft photons contribution (see below).
\section{Scattering channel}
\label{scatt_channel}
Let us consider now the elastic electron muon scattering
\ba
e(p_1) +\mu(p) \to e (p_1')+\mu(p') \nn
\ea
which is the crossed process of (\ref{eq:eq3}).
The Born cross section is
the same for the scattering of electrons and positrons on the same target.
Taking the experimental data from the scattering of electron and positron
on the same target (muon or proton), one can measure the difference of the
corresponding cross-sections which is sensitive to the interference of the
one and two photon exchange amplitudes. For the case of  proton target, in
the Laboratory (Lab) frame, the differential cross section as a function of the
energy of the initial electron, $E$ and of the electron scattering angle,
$\theta_e$, was derived in Ref. \cite{Ro50}:
\ba
\frac{\dd\sigma^{ep}}{\dd\Omega}= \frac{\alpha^{2}}{4E^{2}}
\frac{\cos^{2}\frac{\theta_e}{2}}{\sin^{4}\frac{\theta_e}{2}}
\frac{1}{\rho}\left [\frac{F_{E}^{2}+\tau F_{M}^{2}}{1+\tau}+2\tau
F_{M}^{2}\tan^{2}\frac{\theta_e}{2} \right ], \nonumber \\
\rho=1+\frac{2E}{m}\sin^2\frac{\theta_e}{2}, \quad
\tau=\frac{-t}{4m^{2}}=\frac{E^2}{m^2\rho}\sin^2\frac{\theta_e}{2},
 \label{eq:eq21}
\ea
and it is known as the Rosenbluth formula.
The Sachs electric and magnetic proton form
factors, $F_{E}$ and   $F_{M}$ are related to the
Pauli and Dirac form factors by  $F_{E}=F_{1}-\tau F_{2}$,
$F_{M}=F_{1}+F_{2}$. For the scattering on
muon, one replaces $F_{1}=1$, $F_{2}=0$ and Eq. (\ref{eq:eq21}) becomes
\begin{equation}
\frac{\dd\sigma^{e\mu}_B}{\dd\Omega}=
\frac{\alpha^{2}(s^{2}+u^{2}+2tm^{2})}{2m^{2}\rho^{2}t^{2}},\quad
s=2p_{1}p=2m E,\quad t=-2p_{1}p_{1}',\quad u=-2pp_{1}'=-\frac{s}{\rho}.
\label{eq:eq22}
\end{equation}
The charge-odd contribution to the cross section of $e\mu-$elastic scattering
is:
\begin{eqnarray}
\nonumber \left(\frac{\dd\sigma_{virt}^{e\mu}}{\dd \Omega_{e}}\right)_{odd}&=&
-\frac{\alpha^{3}}{2\pi m^{2}\rho^{2}} Re \left(\mathcal{D}^{sc}\right), \\
\mathcal{D}^{sc} &=&\frac{1}{t}[\mathcal{D}(s,t)-\mathcal{D}(u,t)]
=\frac{2}{t^{2}}[s^{2}+u^{2}+2tm^{2}]
\ln\frac{-u}{s}\ln\frac{-t}{\lambda^{2}}+\mathcal{D}^{sc}_{virt}
 \label{eq:eq23}
\end{eqnarray}
with
\begin{eqnarray}
\mathcal{D}^{sc}_{virt}&=&\frac{s-u}{t}
\left [
\frac{1}{2}\ln^{2}
\left(\frac{-t}{m^{2}} \right )
-\frac{\tau}{1+\tau}\ln \left (\frac{-t}{m^{2}}\right )
+m^{2}\bar{F}_{Q}\left ( 6\tau +2 -\frac{2\tau^{2}}{1+\tau}\right )
\right ]
\nonumber \\
&+&\frac{s}{t}\left [-\ln^{2}\frac{s}{-t} +\pi^{2}
+2\Li{2}{ 1+\frac{m^{2}}{s}}\right ]
-\frac{u}{t}\left[-\ln^{2}\frac{u}{t} +2\Li{2}{
1+\frac{m^{2}}{u}}\right ]
\nonumber \\
&+&\frac{(1-2\tau)}{(-4\tau)}
\left[
2\ln\left (\frac{s}{-u}\right )
\ln\left(\frac{-t}{m^{2}}\right ) +\ln^{2} \left(\frac{-u}{m^2}\right )
-\ln^{2}\left( \frac{s}{m^{2}}\right ) +\pi^{2}\right. \nonumber \\
&+&\left . 2\Li{2}
{1+\frac{m^{2}}{s}}-2\Li{2}{ 1+\frac{m^{2}}{u}}\right ]
+\left (
2m^{2}-\frac{su}{t}\right)
\left[ \frac{\ln\frac{s}{m^{2}}}{m^{2}+s}
-\frac{\ln\frac{-u}{m^{2}}}{m^{2}+u}\right ]
\label{eq:eq24}
\end{eqnarray}
with the help of the following relation:
\begin{equation}
m^2\bar{F}_Q=-\frac{1}{4\sqrt{\tau(1+\tau)}}\left [
\pi^2+\ln(4\tau)\ln x+\Li{2}{-2\sqrt{\tau x}}-\Li{2}{\frac{2\sqrt{\tau}}{\sqrt{x}}}\right ],
\end{equation}
where
$$x=\frac{\sqrt{1+\tau}+\sqrt{\tau}}{\sqrt{1+\tau}-\sqrt{\tau}}.
$$
\section{Soft photon emission}
\label{soft_phot_glav}
In this section the emission of soft real photons in the Lab reference frame for $e\mu$- scattering is calculated. Following Ref. \cite{Ma00}, the odd
part of cross section
\begin{equation}
\frac{\dd\sigma^{soft}}{\dd\sigma_{0}}=-\frac{\alpha}{4\pi^{2}}
\cdot 2\int
\frac{\dd^{3}k}{\omega}\left (\frac{p_{1}'}{p_{1}'k}-\frac{p_{1}}{p_{1}k}\right)
\left (\frac{p'}{p'k}-\frac{p}{pk}\right)_{ S_0,~\omega< \Delta\varepsilon}
 \label{eq:eq25}
\end{equation}
must be calculated in the special reference frame $S_{0},$ where the sum
of the three-momenta of proton and of the recoil proton is zero
$\vec{z}=\vec{k}+\vec{p}'=0.$ Really, in this frame, the on-mass shell
condition of the scattered muon $\delta[(z-k)^{2}-m^{2}]$,
$z=p_{1}+p-p_{1}'$ does not depend on the direction of the emitted photon.
The photon energy can be determined as the difference of the energy of the
scattered electron and the corresponding value for the elastic case: the maximum
value of the photon energy $\Delta \varepsilon$ in
the $S_{0}$ frame is related with the energy of the scattered electron, detected
in the Lab frame $\Delta E$ as (see \cite{Ma00}, Appendix C),
\begin{equation}
\Delta\varepsilon= \rho \Delta E.
\label{eq:eq26}
\end{equation}
The calculation of the soft photon integral with $\omega < \Delta \varepsilon$
can be performed using t'Hooft and M. Veltman approach (see
\cite{HV79}, Section 7). We find
\begin{eqnarray}
\nonumber \frac{\dd\sigma_{e\mu}^{soft}}{\dd\Omega}&=&
-\frac{\alpha^{3}}{\pi}\frac{ (s^{2}+u^{2}+2tm^{2})}
{2m^{2}\rho^{2}t^{2}}\left \{2\ln\rho\ln\left [\frac{(2\rho\Delta E)^2}{\lambda^2 x}\right ]+\mathcal{D}^{sc}_{soft}\right \},
\\
\mathcal{D}^{sc}_{soft}&= &-2\Li{2}{1-\frac{1}{\rho x}}+2\Li{2}{1-\frac{\rho}{x}},~ \rho=\frac{s}{-u},
 \label{eq:eq27}
\end{eqnarray}
which is in agreement with Ref. \cite{Ma00}.

The sum $(\dd\sigma_{e\mu}^{virt}+\dd\sigma_{e\mu}^{soft})_{odd}$  has the form
\ba
\left(\frac{\dd\sigma_{e\mu}^{virt}}{\dd\Omega_{e}}+\frac{\dd\sigma_{e\mu}^{soft}}
{\dd\Omega_{e}}\right)_{odd}
&=&\frac{\alpha^{3}}{2\pi m^{2}\rho^{2}}\frac{(s^{2}+u^{2}+2tm^{2})}{t^{2}}
\left [-2\ln\rho\ln\frac{(2\rho\Delta E)^{2}}{-t x}+\Xi
\right ],
\nonumber \\
\Xi&=& Re \left[-\frac{t^{2}\mathcal{D}^{sc}_{virt}}{s^{2}+u^{2}+2tm^{2}}-\mathcal{D}^{sc}_{soft}\right].
 \label{eq:eq28}
\ea
and it is independent from the photon mass $\lambda$.

\begin{figure}
\includegraphics[scale=1.0]{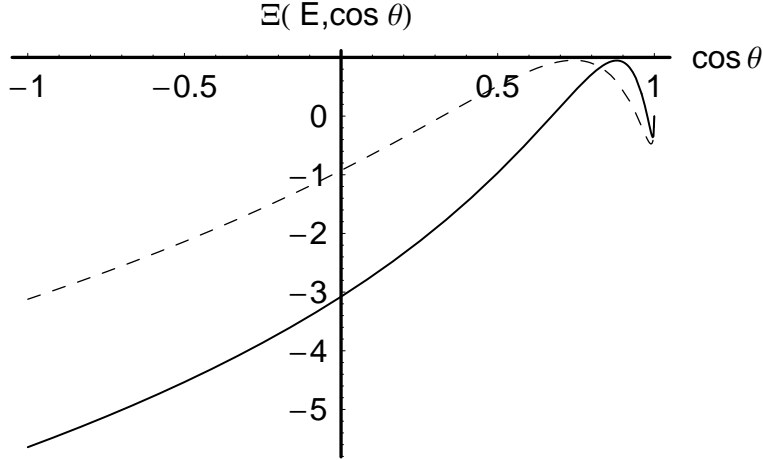}
\caption{$\Xi(s,\cos\theta)$ for $E=5 m$ (dashed line) and $E=10 m$,
$m$ is muon mass.}
\label{fig:Xi}
\end{figure}

The function $\Xi$ is shown in Fig. \ref{fig:Xi} as a function of $\cos\theta_e$ for given $E/m$ .

\begin{figure}
\includegraphics[scale=1]{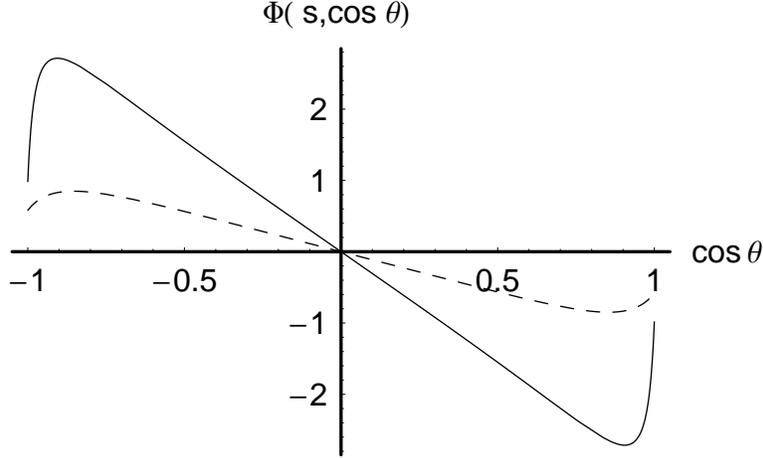}
\caption{$\Phi(s,\cos\theta)$, for $s=10 m^2$ (dashed line) and $s=20 m^2$,
$m$ is muon mass.}
\label{fig:1}
\end{figure}

The ratio between the difference and the sum (corresponding to the Born cross section)
of the cross sections for $e^\pm \mu$ scattering is:
\ba
\frac{
\dd\sigma^{e^-\mu\to e^-\mu(\gamma)}
-\dd\sigma^{e^+\mu\to e^+\mu(\gamma)}
}
{
\dd\sigma^{e^+\mu\to e^+\mu(\gamma)}+
\dd\sigma^{e^-\mu\to e^-\mu(\gamma)}
}
=\frac{\alpha}{\pi}\left[\Xi -
2 \ln \rho \ln \frac{\left(2\rho \Delta E\right)^2}{-t x}\right].
\ea
The odd contributions to the differential cross section for the process
$e^++{e}^-\to \mu^++\mu^-$, due to soft photon emission, has the form:
\begin{equation}
(\dd\sigma_{soft}^{e^+{e}^-\to \mu^+\mu^-(\gamma)})_{odd} =\dd\sigma_{0}
\left(-\frac{\alpha}{4\pi^{2}}\right )2\int
\frac{\dd^{3}k}{\omega}\left(-\frac{p_{-}}{p_{-}k}+\frac{p_{+}}{p_{+}k}\right )
\left(\frac{q_{+}}{q_{+}k}-\frac{q_{-}}{q_{-}k}\right )_{S_{0},\omega<\Delta
\varepsilon}. \label{eq:eq29}
\end{equation}
Again, the integration must be performed in the special frame $S^{0}$,
 where  $\bar{p}_{+}+\bar{p}_{-}-\bar{q}_{+}=\bar{q}_{-}+\bar{k}=0$. In this frame we have
\ba
(q_{-}+k)^{2}-m^{2}&=&2(E_{-}+\omega)\omega \approx 2m \omega =
(p_{+}+p_{-}-q_{+})^{2}-m^{2}=4E(E-\varepsilon_{+}), \nonumber \\
E-\varepsilon_{+}&=&\frac{m}{2E}\Delta\varepsilon.
 \label{eq:eq30}
\ea
In the elastic case $E-\varepsilon_{+}^{el}=0$ and the
photon energy in the Lab system is
\begin{equation}
\Delta E=\varepsilon_{+}^{el}-\varepsilon_{+}=\frac{m}{2E}\Delta \varepsilon.
 \label{eq:eq31}
\end{equation}
The t'Hooft-Veltman procedure for soft photon emission contribution leads to:
\begin{equation}
\frac{\dd\sigma_{ann}^{soft}}{\dd\Omega}=\frac{\dd\sigma_{0}}{\dd\Omega}.
\frac{2\alpha}{\pi}\left[\ln\left(\frac{4E\Delta E}{m
\lambda}\right )^2\ln\frac{1+\beta c}{1-\beta c} +
\mathcal{D}^{ann}_{S}\right ]
 \label{eq:eq32}
\end{equation}
with
\ba
\mathcal{D}^{ann}_{s}&=&\frac{1}{2}\Li{2}{\frac{-2\beta(1+c)}{(1-\beta)(1-\beta c)}}
+\frac{1}{2}\Li{2}{\frac{2\beta(1-c)}{(1+\beta)(1-\beta c)}} \nonumber \\
&&-\frac{1}{2}\Li{2}{\frac{-2\beta(1-c)}{(1-\beta)(1+\beta c)}}
-\frac{1}{2}\Li{2}{\frac{2\beta(1+c)}{(1+\beta)(1+\beta c)}}.
 \label{eq:eq33}
\ea
The total contribution (virtual and soft) is free from infrared
singularities and has the form
\ba
&\frac{\dd\sigma_{ann}}{\dd\Omega}&=\frac{\alpha^{3}\beta}{2\pi
s}(2-\beta^{2}+\beta^{2}c^{2})\Upsilon, \quad
\Upsilon= 2\ln\frac{1+\beta c}{1-\beta c}
\ln\Big(\frac{2\Delta E}{m}\Big)+\Phi(s,\cos\theta), \nonumber \\
&&\Phi(s,\cos\theta)=\mathcal{D}^{ann}_{S}
-\frac{\mathcal{D}^{ann}_{V}}{2-\beta^2+\beta^2 c^2}.
 \label{eq:eq34}
\ea
The quantity $\Phi(s,\cos\theta)$ is presented in Fig. \ref{fig:1}.

The relevant asymmetry can be constructed from (\ref{eq:eq6})
\ba
A=\frac{4\alpha}{\pi}\Upsilon.
\ea

\section{Crossing Symmetry}
\label{CrossingSymmetry}
In this section we formally consider the relations between the kinematical
variables in the scattering and in the annihilation channel,
$e^+ + e^- \to p+\bar p$. The reduced form of the differential elastic $ep$ scattering cross section,
commonly used, is defined as $\sigma_{red}=\tau F^{2}_{M}+\varepsilon F^{2}_{E}$
and it is related to the differential cross section by:
\begin{equation}
\frac{\dd\sigma}{\dd\Omega}=
\sigma_{M}\sigma_{red},~
\sigma_{M}=
\frac{\alpha^{2}\cos^{2}\frac{\theta_{e}}{2}}{4E^{2}\sin^{4}\frac{\theta}{2}}
\frac{1}{\rho\varepsilon(1+\tau)}, \
\varepsilon = \frac{1}{1+2(1+\tau)\tan^{2}\frac{\theta_{e}}{2}}.
\label{eq:eq36}
\end{equation}
where $\varepsilon $ is the polarization of the virtual photon, and varies
from $\varepsilon=0$, for $\theta_{e} = \pi$ to
$\varepsilon=1$, for $ \theta_{e} = 0$.

The crossing relation between the
scattering channel $e+p \to e+p$ and the annihilation channel $e^+ +e^- \to p+\bar{p}$
consists in replacing the variables of the scattering channel $s=2p_1p=2Em$ and  $Q^2=-t$ according to
\begin{equation}
 m^2+s \to t = -2E^{2}(1-\beta c), \ Q^2\to -s=-4E^2,~cos \theta =\cos \hat{\vec{p}_{-}\vec{q}_{-}},
 \label{eq:eq38}
\end{equation}
where $\theta$ is the angle of the antiproton with respect to the incident electron, in the center of mass system (CMS).

The following relation holds, for the annihilation channel:
\begin{equation}
\cos^{2}\theta=\frac{(t-u)^{2}}{s(s-4M^{2})},~s+t+u=2m^2.
 \label{eq:eq39}
\end{equation}
On the other hand, in the scattering channel, one has:
\begin{equation}
\frac{1+\varepsilon}{1-\varepsilon}=
\frac{\cot^{2}\frac{\theta_{e}}{2}}{1+\tau}+1
=\frac{(s-u)^2}{Q^2(Q^2+4M^2)},~Q^2=s+u.
 \label{eq:eq40}
\end{equation}
Therefore one proves the validity of the crossing relation:
\begin{equation}
\cos\theta= \sqrt{\frac{1+\varepsilon}{1-\varepsilon}}\equiv y,
 \label{eq:eq41}
\end{equation}
based on the analytical continuation from the annihilation channel to the scattering one.
This relation was derived in Refs. \cite{Re99,RTG}. Using this relation and the
property of the 2$\gamma$ contribution to the annihilation cross-section
$\left(\frac{\dd\sigma}{\dd\Omega}(\theta)\right)_{2\gamma}=
-\left(\frac{\dd\sigma}{\dd\Omega}(\pi-\theta)\right )_{2\gamma}$ i.e.
$\left(\frac{\dd\sigma}{\dd\Omega}(\theta)\right )_{2\gamma}=
\cos\theta f(\cos^{2}\theta,s)$ the authors of Ref. \cite{RTG} built
the ansatz for $2\gamma$ contribution to $ep$-elastic scattering
\begin{equation}
\frac{\dd\Delta\sigma}{\dd \Omega_e}(e^{-}p\to e^{-}p)= yf(y^{2},Q^{2}); \
f(y^{2},Q^{2})=c_{0}(Q^{2})+y^{2}c_{1}(Q^{2})+y^{4}c_{2}(Q^{2})+\ldots .
 \label{eq:eq42}
\end{equation}
This property follows from the change of the sign of the contribution for virtual and
real photon emission when the ($s\leftrightarrow u$) transformation is applied
(see Eqs. (\ref{eq:eq24},\ref{eq:eq27},\ref{eq:eq28}) and relation (\ref{eq:eq40})).

This form of the contribution of the $1\gamma\bigotimes 2\gamma$ interference to
the differential cross section derives explicitly from C-invariance and crossing symmetry
of electromagnetic interactions and excludes any linear function of $\varepsilon$ for a
possible parameterization of such contribution.

Let us  note that not only the elastic channel must be taken into account:
the interference of the amplitudes corresponding to the emission of a photon by
electron and by proton must be considered, too.

Evidently, the relations derived above are valid for the considered processes
with electrons and heavy lepton setting respectively $F_E=F_M=1$.

\section{Derivation of the additional structure: annihilation channel}
\label{DerivationOfF3}
Let us start from the following form of the matrix element for the process
$e^+(p_+)+e^-(p_-)\to \mu^+(q_+)+\mu^-(q_-)$ in presence of $2\gamma$ exchange:
\begin{eqnarray}
M_{2} =\frac{ i\alpha^2}{s} \bar{v}(p_{+}) \gamma_{\mu} u(p_{-})\times
\bar{u}(q_{-})
\left (G_{1}\gamma_{\mu}-\frac{G_{2}}{m}\gamma_{\mu}\hat{P}+4\frac{1}{s}G_3\hat{\Delta}Q_\mu\right ) v(q_{+}),
\label{eq:eq44}
\end{eqnarray}
where the amplitudes $G_{i}$ are complex functions of the two kinematical variables $s$, and $t$.

To calculate the structure $G_{3}$ from the $2\gamma$ amplitude (see Eq. (\ref{eq:eq8})),
 both Feynman diagrams (Figs. \ref{Fig:fig1}a and \ref{Fig:fig1}b) must be taken
into account. Similarly to Section II, only one of them can be calculated explicitly (the uncrossed one), whereas
the other can be obtained from this one by appropriate replacements.

To extract the structure $G_3$ we multiply Eq. (\ref{eq:eq44}) subsequently by
\ba
&&\bar{u}(p_-)\gamma_\lambda v(p_+)\times \bar{v}(q_+)\gamma_\lambda u(q_-), \nonumber \\
&&\bar{u}(p_-)\hat{Q} v(p_+)\times \bar{v}(q_+) u(q_-), \label{eq:terms} \\
&&\bar{u}(p_-)\hat{Q} v(p_+)\times \bar{v}(q_+)\hat{\Delta} u(q_-),
\nonumber
\ea
and perform the summation on fermions spin states.

Solving the algebraical set of equations we find
\ba
G^a_{1}&=&\frac{1}{\beta^4\sin^{4}\theta} \left \{
(8 B^a + A^a\beta^2\sin^{2}\theta)(1-\beta^2\cos^2\theta)-4C^a\beta\cos\theta
\left [2-\beta^2(1+\cos^{2}\theta)\right ]\right \},
\nonumber \\
G^a_{2}&=&\frac{1}{\beta^4\sin^{4}\theta} \left\{
\beta(1-\beta^2)(A^a \beta\sin^{2}\theta -8C^a \cos\theta)  +4 B^a
\left [2-\beta^2(1+\cos^2\theta)\right ]\right \}
 \label{eq:eq46} \\
G^a_{3}&=&\frac{1}{\beta^3\sin^{4}\theta} \left [
-A^a\beta^{2}\sin^{2}\theta\cos\theta-8 B^a\cos\theta+4\beta C^a (1+\cos^{2}\theta)\right ],
\nonumber
\ea
with
\ba
A^a&=&\int\frac{\dd^4k}{i\pi^2}\frac{1}{(\Delta)(Q)(P_+)(P_-)}\frac{1}{s} Tr
(\hat{p}_{+}Z\hat{p}_{-}\gamma_{\lambda})
\times\frac{1}{4}Tr\left [(q_{-}+m)T(\hat{q}_{+}-m)\gamma_{\lambda}\right ],
\nonumber\\
B^a&=&\int\frac{\dd^4k}{i\pi^2}\frac{1}{(\Delta)(Q)(P_+)(P_-)}  \frac{m}{s^2}
Tr(\hat{p}_{+}Z\hat{p}_{-} \hat{Q})\times\frac{1}{4}Tr\left [ (\hat{q}_{-}+m)T(\hat{q}_{+}-m)\right ],
 \label{eq:eq48}\\
C^a&=&\int\frac{\dd^4k}{i\pi^2}\frac{1}{(\Delta)(Q)(P_+)(P_-)}  \frac{1}{s^2} Tr
(\hat{p}_{+}Z\hat{p}_{-}\hat{Q})\times\frac{1}{4}Tr\left [ (\hat{q}_{-}+m)T(\hat{q}_{+}-m)\hat{\Delta}\right ].
 \nonumber
\ea
The explicit value for $G_3^a$ is:
\ba
G^a_{3}&=&\frac{2 s}{\beta^3(1-c^2)^{2}}
\left \{
 \frac{1}{2} G_Q(1-c^2)\beta^3(1 - \beta c )\right .
\nonumber \\
&+&\frac{1}{2}H_Q\beta^2(1-c^2)\left [  c(-3+ 5\beta^2)- \beta - \beta c^2\right ]
\nonumber \\
&+& F_\Delta c\left [ 1- 4\beta^2+2\beta^4+c^2\beta^2(3-4\beta^2)
                - 2\beta c(1-2\beta^2)\right ]
\label{eq:eq49} \\
&+&F_Q\beta\left [ -c^2 + \beta c \left (-\frac{1}{2}- 4\beta^2 c^2+\frac{5}{2}c^2\right )
                  + \beta^2 \left (-\frac{1}{2} + 2\beta^2 c^2 +
           \frac{3}{2}c^2\right)\right ]
\nonumber \\
&-&2 J s \beta^2c(1-c^2)(1-\beta^2)(1  - \beta c)
\nonumber \\
&+&\left .F c \left [1+ \beta^2 c^2-2\beta^4-4\beta^4c^2
              +\beta c(-3 + 4\beta^2 + 2\beta^4
              +\beta^2c^2)\right ]
\right  \}.
\nonumber
\ea
The contributions from the crossed Feynman diagram can be obtained from Eqs. \ref{eq:eq49} by:
\begin{equation}
(A^b,B^b,C^b)_{crossed}=-[A^a,B^a,C^a(\cos\theta\to-\cos\theta)]_{uncrossed}.
\end{equation}
As one can see,  in the quantities $G_1$, $G_2$, and $G_3$  infrared divergencies are present.

\section{One spin asymmetry}
\label{ProtonSpinAsymmetry}
Let us consider now the process of electron interaction with a  heavy lepton (point-like proton).
For clearness the expressions are written for the proton case. For the  interaction of electrons with heavy leptons, $\mu$ or $\tau$,
one should take $G_E=G_M=1$ and use the relevant mass replacement.

The target spin asymmetry for heavy fermions production process
$e^+(p_+)+e^-(p_-) \to p(q_+)+\bar{p}(q_-)$ (in CMS frame)  is defined as
\ba
\frac{\dd\sigma^{\uparrow}-\dd\sigma^{\downarrow}}{\dd\sigma^{\uparrow}+\dd\sigma^{\downarrow}}=(\vec{a}\vec{n})R_n,
\label{AssymetryDefinition}
\ea
where $\vec{a}$ is the proton polarization vector,
$\vec{n} = \left(\vec q_- \times \vec p_-\right)/\left|\vec q_- \times \vec p_-\right|$ is the unit vector normal to
the scattering plane,
$\dd\sigma^{\uparrow}$ is the cross section of processes with proton
polarization vector $\vec a$, $\dd\sigma^{\downarrow}$ is the cross section of
processes with proton
polarization vector  $-\vec a$.
Thus the denominator of the left hand side in Eq. (\ref{AssymetryDefinition}) is the unpolarized cross section of process
$e^+e^- \to p +\bar p$ which is well-known \cite{IoffeLipatovKhoze}
(for the case of proton-antiproton creation):
\ba
    \frac{\dd\sigma^{e^+e^- \to p \bar p}}{d \Omega}
    &=&
    \frac{\alpha^2 \beta}{4 s}
    \left[
        \left(1+\cos^2\theta\right) \left|G_M\right|^2
        +
        \left(1-\beta^2\right) \left|G_E\right|^2 \sin^2 \theta
    \right],
    \label{eeToPPBorn}
\ea
with $\beta=\sqrt{1-\frac{4M^2}{s}}$ is the velocity of proton in c.m. frame, $s$ is
the total energy square and $\theta$ is the angle between vectors $\vec q_-$ and $\vec p_-$.

The difference of cross sections in (\ref{AssymetryDefinition}) is originated by the $s$-channel discontinuity of interference of the Born-amplitude
with TPE amplitude
\ba
    \dd\sigma^{\uparrow}-\dd\sigma^{\downarrow} \sim
   Re \sum \left(A_{elastic}^+ \cdot A_{TPE} + A_{elastic} \cdot A_{TPE}^+\right).
\ea
Using the density matrix of final proton
$u(p) \bar u(p) = (\hat{p}+M)(1-\gamma_5 \hat a)$ one gets
\ba
    && Re\sum \left(A_{elastic}^+ \cdot A_{TPE} + A_{elastic} \cdot A_{TPE}^+\right) =
   32\frac{\left(4\pi\alpha\right)^3 \left(2\pi i\right)^2}{s \pi^2}
    Re(Y), \nonumber \\
  &&Y=
\int \frac{dk}{i \pi^2}
    \frac{1}{(\Delta)(Q)(+)(-)}
    \times
    \frac{1}{4} Tr\left[
        \hat p_1 \gamma^\alpha \hat p_1' \gamma^\mu
        \left(\hat k - \hat \Delta\right)
        \gamma^\nu
    \right]\times
\label{SpinAssymetrySpurs} \\
    &&\times
    \frac{1}{4} Tr\left[
        \left(\hat p - M\right)
        \left(- \gamma_5 \hat a\right)
        \gamma_\alpha
        \left(\hat p' + M\right)
        \gamma_\nu
        \left(\hat k - \hat Q + M\right)
        \gamma_\mu
    \right].
    \nonumber
\ea
Performing the loop-momenta integration the right hand side  of
Eq. (\ref{SpinAssymetrySpurs}) can be expressed in terms of
basic integrals (see Appendix B)
\ba
    Re(Y) = 4 M (a, \Delta, Q, P) Im\left(F_Q - G_Q + H_Q\right),
    \label{ImPartDefinition}
\ea
where $(a, \Delta, Q, P) \equiv \varepsilon^{\mu\nu\rho\sigma} a_\mu \Delta_\nu Q_\rho
P_\sigma= (\sqrt{s}/2)^3 (\vec a \vec n) \beta \sin\theta$.
Using the expressions listed in Appendix B we have:
\ba
    Im\left(F_Q - G_Q + H_Q\right) =\frac{\pi}{s}\psi(\beta)=
    \frac{\pi}{s \beta^2}
    \left(\frac{1-\beta^2}{\beta}
    \ln\frac{1+\beta}{1-\beta} - 2\right). \label{AnnihilImPart}
\ea

\begin{figure}
\begin{center}
\includegraphics[width=12cm]{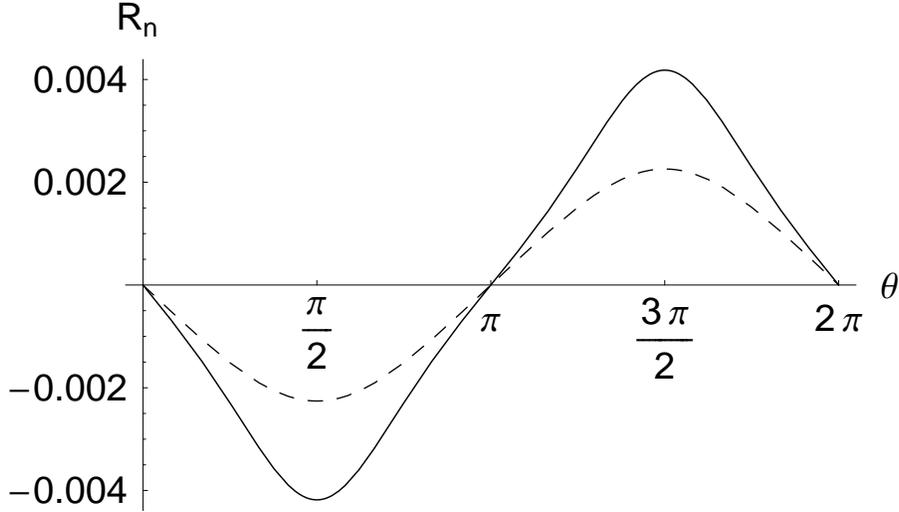}
\caption{\label{polarpoint}  Asymmetry $R_n$
for the case of structureless proton for energies $s=5$ GeV$^2$ (dashed line) and $s=15$ GeV$^2$ (solid line).}
\end{center}
\end{figure}

Thus, after standard algebra, the following expression for
spin asymmetry can be obtained for the processes
$e^++e^-\to \vec{p}+\bar{p}$:
\ba
    R_n =
    2\alpha\frac{M}{\sqrt{s}}~
    \frac{\beta\psi(\beta)\sin\theta}
        {2 - \beta^2 \sin^2 \theta
    }.
\ea
and it is shown in Fig. \ref{polarpoint}, as a function of $\theta$
at several values of $s$ for the case of structureless proton.

\begin{figure}
\begin{center}
\includegraphics[width=12cm]{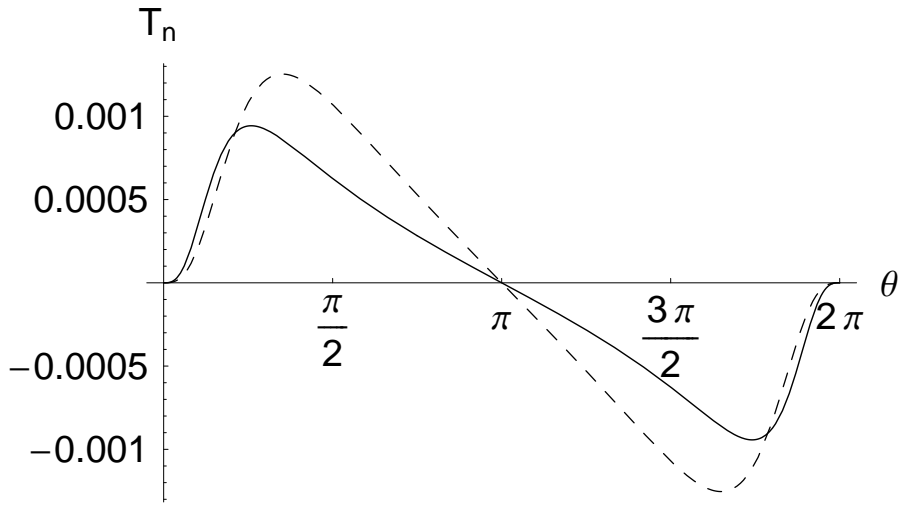}
\caption{\label{polareppoint} Asymmetry $T_n$ in case of structureless
proton for  $E=5$ GeV (dashed line) and $E=10$ GeV (solid line).}
\end{center}
\end{figure}

Such considerations apply to the scattering channel when the
initial protons is polarized.
Similarly to (\ref{AnnihilImPart}) one finds
\ba
    Im_s\left(\bar F_Q - \bar G_Q + \bar H_Q\right) =
    -\frac{\pi}{s + M^2}. \label{ScatteringImPart}
\ea
(note that the $s$-channel imaginary part vanishes for the crossed photon diagram amplitude).
The contribution of the polarization vector appears in the same combination
\ba
(a, \Delta, Q, P) = \frac{1}{2} (a, p, p_1, q) = \frac{M E^2}{2 \rho} \sin\theta (\vec a \vec n).
\ea
The single spin asymmetry for the process
$e^- + \vec{p} \to e^- +p$ (the initial proton is polarized) has the form:
\ba
\frac{\dd\sigma^{\uparrow}-\dd\sigma^{\downarrow}}{\dd\sigma^{\uparrow}+\dd\sigma^{\downarrow}}=(\vec{a}\vec{n}) T_n
\ea
with
\ba
    T_n =
    \frac{\alpha}{2 M^2}\frac{s^2}{s+M^2}
    \left(1+ \tau\right)
    \frac{\epsilon}{\rho \sigma_{red}}
    \sin\theta
    \tan^2\frac{\theta}{2}.
\ea
This quantity  is shown in Fig. \ref{polareppoint}
for the case of structureless proton
as a function of $\theta$,  for two values of $s$.
The  asymmetry decreases when the c.m.s. energy growth, so on experiment it
is useful to measure the asymmetry near the treshold of proton (or heavy lepton) production.

\section{Conclusions}

We calculated QED radiative corrections to the differential cross-section of the processes
$e^++{e}^-\to \mu^++{\mu}^-(\gamma)$, $e^{\pm}+\mu\to e^{\pm}+\mu (\gamma)$, arising from the interference between the Born and the box-type amplitudes. The relevant part of soft photon emission
contribution, which eliminates the infrared singularities, was also considered.

According to the previous discussion, the box Feynman diagram,
calculated in QED frame, can also be applied to lepton-hadron interaction.
In the present work, we considered structureless Born approximation
for annihilation and scattering channels. A realistic case with
Born and box diagram depending on form-factors will be considered further
(in preparation).

Angular asymmetry, charge asymmetry as well as target spin asymmetry were calculated. These quantities are free from infrared and electron mass singularities. Numerical applications
show that these observables are large enough to be measured (see Figs. \ref{fig:Xi}-\ref{polareppoint}).

The charge-asymmetry properties of radiative corrections in the
annihilation channel induce non-trivial terms in the cross
section due to crossing symmetry relations (see (\ref{eq:eq42})).

The parametrization (\ref{eq:eq44}) for the contribution to the matrix
element arising from box-type
diagrams in terms of three additional functions
$G_i(s, t)$, $i = 1, 2, 3$ suffers from infrared divergencies, which can be eliminated by taking into account soft photon emission expressed
in terms of structures $G_1$, $G_2$, $G_3$. This procedure results in
replacing $\ln (m/\lambda)$ with $\ln\Delta$.

The results obtained here, for the processes  $e^\pm p\to e^\pm p(\gamma)$ are particularly interesting in view of the experiments planned at Novosibirsk \cite{Ni05} and at JLab \cite{Pe05} as well as $e^++e^-\to N\bar{N} (\gamma)$, which can be investigated at Frascati \cite{Mi05} and Bejing \cite{Li04}.

\section{Appendix A: Trace calculation.}

The explicit expressions for the polynomials $P_i$ are:
\ba
P_1&=&8\{-(\Delta^2)^3-\Delta^2\sigma^2+2\sigma^3+[
(\Delta^2)^2-2\Delta^2\sigma ] m^2\},
\nonumber \\
P_2&=&16[(\Delta^2)^2\sigma-\sigma^3+\Delta^2\sigma m^2], \nonumber \\
P_3&=&8(\{2(\Delta^2)^2\sigma-2\sigma^3+m^2[
(\Delta^2)^2+2\Delta^2\sigma-\sigma^2]+\Delta^2m^4\},
\nonumber \\
P_4&=&8[5(\Delta^2)^2-6\Delta^2\sigma+5\sigma^2-5\Delta^2m^2], \nonumber \\
P_5&=&8[(\Delta^2)^3-2(\Delta^2)^2\sigma+\Delta^2(\sigma^2-m^2\Delta^2)], \nonumber \\
P_6&=&8\{(\Delta^2)^3-2(\Delta^2)^2\sigma+\Delta^2\sigma^2+m^2[-(\Delta^2)^2-2\Delta^2\sigma+2\sigma^2]-2\Delta^2m^4\},
\nonumber \\
P_7&=&8[-(\Delta^2)^3+2(\Delta^2)^2\sigma-\Delta^2\sigma^2+(\Delta^2)^2m^2], \nonumber \\
P_8&=&8\{-(\Delta^2)^3+(\Delta^2)^2\sigma-3\Delta^2\sigma^2+3\sigma^3-m^2[(\Delta^2)^2+3\Delta^2\sigma]\}.
\ea

\section{Appendix B: Useful integrals.}
\label{AppendixWithIntegrals}

In the calculation of $ep$ scattering we use the following set
of scalar integrals with three and four denominators \cite{KM76a}.
\ba
F_\Delta&=&\frac{-i}{\pi^2}\int\frac{\dd^4k}{(\Delta)(P_+)(P_-)}=\frac{1}{s}
\left [\frac{\pi^2}{6}
+\frac{1}{2}\ln^2\frac{s}{m_e^2}\right ], \nonumber \\
F_Q&=&\frac{-i}{\pi^2}\int\frac{\dd^4k}{(Q)(P_+)(P_-)}
\nonumber \\
&=&\frac{1}{s\beta} \left [
\frac{1}{2}\ln^2\frac{1-\beta}{2}-\frac{1}{2}\ln^2\frac{1+\beta}{2}
+\Li{2}{\frac{1+\beta}{2}}-\Li{2}{\frac{1-\beta}{2}}\right ],
\nonumber \\
H&=&\frac{-i}{\pi^2}\int\frac{\dd^4k}{(\Delta)(Q)(P_+)}
=G=\frac{-i}{\pi^2}\int\frac{\dd^4k}{(\Delta)(Q)(P_-)}
=-\frac{1}{2(m^2-t)}\left [
\ln^2\frac{m^2-t}{m^2} \right .\nonumber\\
&+&\left . \left ( 2\ln\frac{m^2-t}{m^2}+\ln\frac{m^2}{m_e^2}\right ) \ln\frac{m^2}{\lambda^2}
-\frac{1}{2}\ln^2\frac{m^2}{m_e^2}-2\Li{2}{-\frac{t}{m^2-t}}
\right ],
\nonumber \\
F&=&\frac{1}{2}s J-G=-\frac{1}{2(m^2-t)}\left[ \left
(2\ln\frac{m^2-t}{m^2}+\ln\frac{m^2}{m_e^2}\right )\ln\frac{s}{m^2} \right .\nonumber \\
&-&\left . \ln^2\frac{m^2-t}{m^2}+\frac{1}{2}\ln^2\frac{m^2}{m_e^2}+2\Li{2}{-\frac{t}{m^2-t}}
\right ],
\nonumber \\
J&=&\frac{-i}{\pi^2}\int\frac{\dd^4k}{(\Delta)(Q)(P_+)(P_-)}=-\frac{1}{s(m^2-t)}
\left [
\left(2\ln\frac{m^2-t}{m^2}+\ln\frac{m^2}{m_e^2}\right )\ln\frac{s}{\lambda^2}
\right ].\label{eq:eq64}
\ea
The terms proportional to $m_e^2/s$, $m_e^2/m_\mu^2$ were neglected. Notations follow Ref.  (\ref{eq:design}).

The vector integrals with  three denominators are:
\ba
&&\frac{1}{i\pi}\int\frac{k^\mu\dd^4k}{(\Delta)(Q)(P_+)}=H_P P^\mu+H_\Delta \Delta^\mu+H_Q Q^\mu, \quad
H_Q=\frac{1}{t}\ln\frac{m^2-t}{m^2},\nonumber \\
&&H_\Delta=\frac{1}{m^2-t}\left (-\ln\frac{m^2}{m^2_e}-\frac{m^2+t}{t}\ln\frac{m^2-t}{m^2}\right ), \nonumber \\
&&H_P=H+\frac{1}{m^2-t}\left (\ln\frac{m^2}{m^2_e}+2\ln\frac{m^2-t}{m^2}
\right ) ,
\label{eq:eq65}\\
&&\frac{1}{i\pi}\int\frac{k^\mu\dd^4k}{(\Delta)(P_+)(P_-)}=G_\Delta \Delta^\mu, \quad
G_\Delta=\frac{1}{s}\left (-2\ln\frac{s}{m^2_e}+\frac{1}{2}\ln^2\frac{s}{m^2_e}+\frac{\pi^2}{6}\right ),
\nonumber \\
&&\frac{1}{i\pi}\int\frac{k^\mu \dd^4k}{(Q)(P_+)(P_-)}=G_Q Q^\mu, \quad
G_Q=\frac{1}{s-4m^2}\left (-2\ln\frac{s}{m^2}+sF_Q\right ).
\nonumber
\ea
Four denominator vector and tensor integrals were defined in (\ref{eq:eq14}).
The relevant coefficients are:
\ba
J_\Delta&=&\frac{1}{2d}\biggl[(F+F_\Delta)\sigma-Q^2(F+F_Q)\biggr], \nonumber \\
J_Q&=&\frac{1}{2d}\biggl[(F+F_Q)\sigma-\Delta^2(F+F_\Delta)\biggr], \quad
F=\frac{1}{2}s J-G,\quad d=\Delta^2Q^2-\sigma^2.
\nonumber \\
K_0&=&-\frac{1}{2\sigma}\biggl[\sigma(F-G+H_P+H_\Delta+H_Q)+H_\Delta(\sigma-\Delta^2)-H_Q(\sigma-Q^2)
\nonumber \\
&&+2P^2J_\Delta(\Delta^2-2\sigma)+\Delta^2G_\Delta-Q^2G_Q-2p^2Q^2J_Q \biggr],
\nonumber \\
K_\Delta&=&-\frac{1}{2\sigma d}\biggl[Q^2\sigma(G-F-H_P-3H_\Delta+6P^2J_\Delta)
\label{eq:eq66}\\
&&+(\Delta^2Q^2+\sigma^2)(H_\Delta-2P^2J_\Delta-G_\Delta)
-Q^4(H_Q-2P^2J_Q-G_Q)\biggr],
\nonumber \\
K_P&=&\frac{1}{2P^2\sigma}\biggl[
2\sigma(H_\Delta-2P^2J_\Delta+H_p+\frac{1}{2}F-\frac{1}{2}G)+Q^2(H_Q-2P^2J_Q-G_Q)
\nonumber \\
&&-\Delta^2(H_\Delta-2P^2J_\Delta-G_\Delta)\biggr],
\nonumber \\
K_Q&=&-\frac{1}{2\sigma d}\bigg[-\Delta^2\sigma A_P+2\Delta^4A_\Delta+(\sigma^2-2\Delta^2Q^2)A_Q\biggr],
\nonumber \\
K_x&=&-\frac{1}{2 d}\left (\sigma A_P+Q^2A_Q-2\Delta^2 A_\Delta\right ),
\nonumber
\ea
where we used
\begin{equation}
A_\Delta=H_\Delta+2\Delta^2J_\Delta-G_\Delta,~
A_Q=H_Q+2\Delta^2J_Q-G_Q, ~
A_P=F-G+H_P+3H_\Delta+6\Delta^2J_\Delta.
\end{equation}

\end{document}